\documentclass[journal=ancac3,manuscript=article,layout=onecolumn]{achemso}
\usepackage{pdfpages,textcomp,gensymb,amsmath,xcolor,multirow,natbib,lineno,physics,array,comment}
\newcommand*\2{$_2$}
\newcommand{\txr}{\textcolor{purple}}
\newcommand{\txb}{\textcolor{black}}
\newcommand{\unsim}{{\sim}}
\newcolumntype{R}[1]{>{\raggedleft\arraybackslash}m{#1}}  

\usepackage{verbatim}

\author{Nithin Abraham}
\email{nithina@iisc.ac.in}
\affiliation[IISc]
{Department of Electrical Communication Engineering, Indian Institute of Science, Bangalore 560012, India}
\author{Kenji Watanabe}
\email{WATANABE.Kenji.AML@nims.go.jp}
\affiliation[NIMS1]
{Research Center for Electronic and Optical Materials, National Institute for Materials Science, 1-1 Namiki, Tsukuba 305-0044, Japan}
\author{Takashi Taniguchi}
\email{TANIGUCHI.Takashi@nims.go.jp}
\affiliation[NIMS2]
{Research Center for Materials Nanoarchitectonics, National Institute for Materials Science,  1-1 Namiki, Tsukuba 305-0044, Japan}
\author{Kausik Majumdar}
\email{kausikm@iisc.ac.in}
\phone{+91-80-2293-2742}
\affiliation[IISc]
{Department of Electrical Communication Engineering, Indian Institute of Science, Bangalore 560012, India}

\title{Room Temperature Single Photon Detection at 1550 nm using van der Waals Heterojunction}

\begin{document}

\newpage
\begin{abstract}
Single-photon detectors (SPDs) are crucial in applications ranging from space, biological imaging to quantum communication and information processing. The SPDs that operate at room temperature are of particular interest to broader application space as the energy overhead introduced by the cryogenic cooling can be avoided. Although silicon-based single photon avalanche diodes (SPADs) are well matured and operate at room temperature, the bandgap limitation restricts their operation at telecommunication wavelength (1550 nm) and beyond. InGaAs-based SPADs, on the other hand, are sensitive to 1550 nm photons, but suffer from relatively lower efficiency, high dark count rate, afterpulsing probability, and pose hazards to the environment from the fabrication process. In this work, we demonstrate how we can leverage the properties of nanomaterials to address these challenges and realise a room temperature single-photon detector capable of operating at 1550 nm. We achieve this by coupling a low bandgap ($\unsim 350~meV$) absorber (black phosphorus) to a sensitive van der Waals probe that is capable of detecting discrete electron fluctuation. We optimize the device for operation at $1550~nm$ and demonstrate an overall quantum efficiency of $21.4\%$ (estimated as $42.8\%$ for polarized light), and a minimum dark count of $\unsim 720~Hz$ at room temperature.
\end{abstract}

\newpage

A single photon around $1550~nm$ wavelength carries a tiny amount of energy, on the order of $10^{-19}~J$, making reliable detection of individual photons highly challenging. Nevertheless, decades of research have enabled us to realize different single-photon detector (SPD) technologies, such as photomultiplier tubes (PMT)\cite{Kume88}, single-photon avalanche detectors (SPAD)\cite{Liu22}, visible light photon counters (VLPC)\cite{Baek10}, superconducting nanowire single-photon detectors (SNSPD)\cite{Marsili13,Pan22,Hu20,Chang21}, transition-edge sensors (TES)\cite{Fukuda11,Lita08}, up-conversion detectors\cite{Bai17}, and quantum dot-based detectors\cite{Weng15}. SPDs play a crucial role in bioimaging, sensing, spectroscopy, astronomy, and also in emerging technologies such as quantum computation and quantum communication. Photonic qubit is a promising platform for realizing these quantum applications as they provide a clear advantage in many aspects, such as room-temperature operation, long-range transmission, low de-coherence, and ease of manipulation.\cite{Slussarenko19,Nielsen04} Successful implementation of these technologies relies heavily on highly sensitive SPDs.

For applications utilizing the visible range of the spectra, Si-based SPADs are the preferred choice due to their superior performance\cite{Thomas10}. However, the large bandgap of silicon limits the operation of these SPADs to wavelengths shorter than $\unsim1.1~\mu m$. The telecom wavelengths around $\unsim1550~nm$, where silica optical fibers exhibit very low attenuation falls outside operating range of Si-SPAD. Operation at these wavelengths is critical for applications such as quantum key distribution (QKD)\cite{Chen21}, which require the transmission of photonic qubits over long distances. The infrared (IR) operation is also essential in other fields, such as astronomy, where infrared imaging detects faint galaxies. InGaAs-SPADs\cite{com1} can extend the operating range to $\unsim1700~nm$, but suffer from relatively lower efficiency than Si-SPADs, need for moderate cooling, environmental hazards from the fabrication of As-based compounds, high dark counts, and high afterpulsing probability\cite{com1,com2,com3}. There are active researches toward improving their performance, especially to improve efficiency, to reduce dark counts with the help of active cooling and to minimize afterpulsing\cite{xu2023}. Superconducting transition-based SPDs are excellent alternatives featuring very high efficiency, ultrafast response, high count rates and broad spectral response, and are presently the preferred choice for high sensitivity detection of IR photons. But their dependence on cryogenic temperature ($<4~K$ for SNSPD\cite{Marsili13} and $\unsim100~mK$ for TES\cite{Fukuda11}) severely limits their application space due to high cost and cooling requirement (for example, space-based applications and large scale deployment of QKD network). A representative summary of the various technologies present for single photon detection at $1550~nm$ is given in Supplementary Table S1.\txb{ The commercially available solutions using these technologies are not yet operating at room tempreature - InGaAs/InP\cite{com1} with efficiency of 32\%@ 225K and SNSPD \cite{com4} with efficiency greater than 90\%@ 2.5K.} The pressing need for an alternate technology that is capable of detecting single photons of wavelength $1550~nm$ and beyond at room temperature with high sensitivity and low dark counts is evident. In this manuscript, we show that the untapped potential of nanomaterial based systems can be a powerful tool to tackle many of these challenges and demonstrate a different approach towards single photon detection. To achieve this, we utilize a low bandgap absorber - black phosphorus (BP) - and couple it with a van der Waals heterostructure designed to detect single electron fluctuations. The fabricated devices are then rigorously tested and verified to detect single photons at $1550~nm$ wavelength with promising efficiency while maintaining room-temperature operation and low dark counts.

\section{Results and Discussion}
\subsection{Device design}
\begin{figure}[ht]
  \centering
  \includegraphics[width=12.14cm]{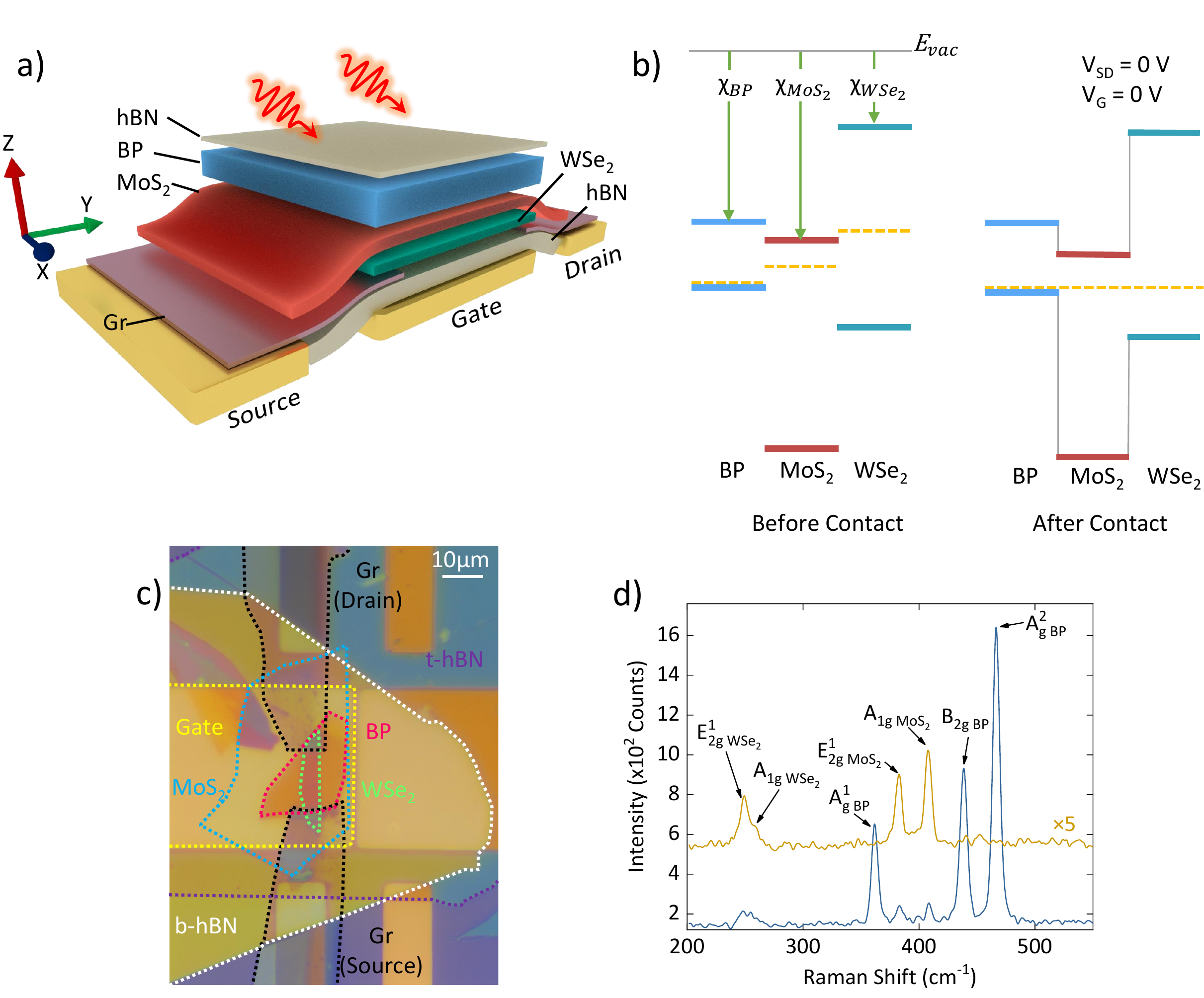}	
  \caption{\label{fig1}\textbf{Device structure and basic characterization.} \textbf{a)} Schematic representation of the device architecture. Few layer flakes of hBN, Gr, WSe\2, MoS\2 and BP are used. The region of overlap between BP with MoS\2 layers in the X-Y plane defines the active area. \textbf{b)} Energy band alignment along the Z-direction before contact (left) and after contact(right) showing the quantum well formation in MoS\2. The orange dashed line indicates the Fermi level. \textbf{c)} Optical image of a fabricated device outlining the different layers. The scale bar is $10~\mu m$. \textbf{d)} Raman characterization of the flakes under a $532~nm$ CW laser illumination. The signal for MoS\2 and WSe\2 is collected from a region without BP on top to avoid the strong absorption in BP.}
\end{figure}

A schematic representation of the proposed design is shown in Figure \ref{fig1}a. The absorbing region ($\unsim 25~nm$-thick BP), and a $\unsim 5~nm$-thick MoS\2 are stacked on a p-type WSe\2 ($\unsim5~nm$ thick) region\cite{Wang18}. MoS\2 forms a type II heterostructure with both BP and WSe\2 (Figure \ref{fig1}b), and hence acts as an electron quantum well. The WSe\2/MoS\2/BP stack forms the channel of a field effect transistor (FET). Few layer Gr flakes are used as the source and drain contacts. The Gr flakes are not in direct contact with BP. A bottom metal pad and bottom hBN forms the gate and gate dielectric of the FET respectively. The entire stack is then capped with a top hBN layer. The details of the fabrication process are given in \textbf{Methods}. A representative optical image of a fabricated device is shown in Figure \ref{fig1}c. Layer identification is confirmed by Raman characterization (Figure \ref{fig1}d). A false-colored high-resolution transmission electron microscopy image of the cross-section of a typical device is shown in Supplementary Figure S1. Repeatable observations were made from four such devices and are discussed in this manuscript. Following a DC analysis (Supplementary Figure S2, S3), the gate voltage ($V_G$) is set at -4V for the results shown in the rest of the manuscript. This is done to deplete MoS\2 of electrons and to make WSe\2 channel and BP channel p-type.

\subsection{Outline of the detection process}
\begin{figure}[H]
  \centering
  \includegraphics[width=14.86cm]{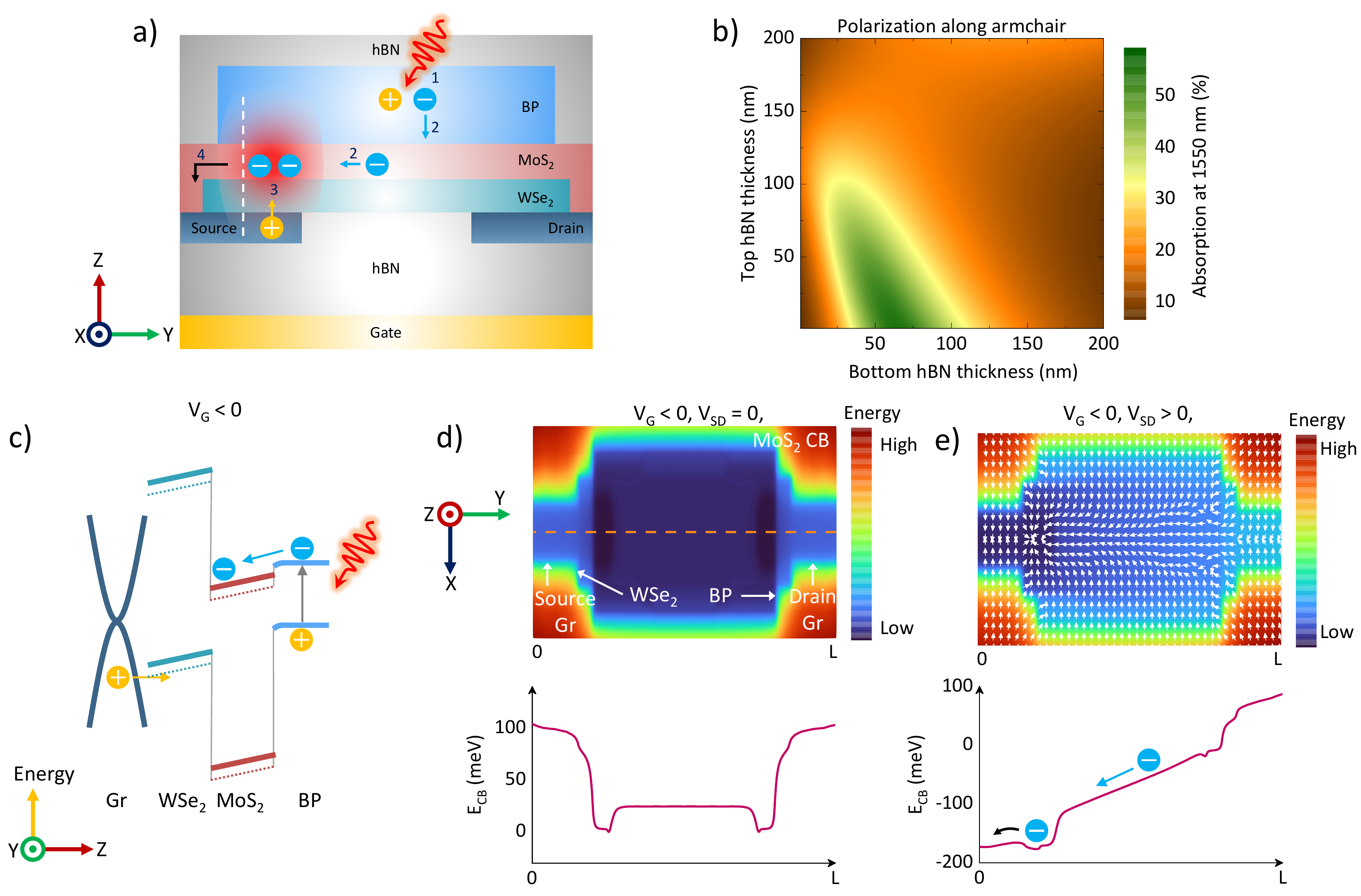}	
  \caption{\label{fig2}\textbf{Principle of operation.} \textbf{a)} A schematic representation showing the photo-generation of electron-hole pair in BP and the dynamics of the carriers in a cross-section of the device along the Y-Z plane. Yellow (blue) arrows indicate the transport of holes (electrons). The black arrow indicates the pumping out of electrons from the island, resetting the detector. \textbf{b)} Results from transfer matrix method-based estimation showing the dependence of the absorption of photons at $1550~nm$ wavelength in BP on the thickness of the top and bottom hBN regions. The thickness of the BP region was fixed at $25~nm$ and was fixed at $5~nm$ for both WSe\2 and MoS\2 regions. Polarization of light was assumed to be aligned with the armchair direction in BP. \textbf{c)} Modulation of the energy bands along the white dashed line shown in \textbf{(a)} and thereby the modulation of hole injection efficiency with the trapping of an electron in MoS\2. Dashed (solid) lines indicate the energy bands before (after) electron trapping. \textbf{d)} Simulated potential energy profile (top panel) of the MoS\2 CB in the X-Y plane under a negative $V_G$ and at $V_{SD}=0$. A rectangular geometry is assumed for all the flakes. Regions of MoS\2 that exhibit overlap with other layers are annotated by labeling the corresponding layer. The formation of the island is shown as the lowest energy region (dark blue) where MoS\2 exhibits overlap with Gr, WSe\2, and BP. The CB energy profile along the orange dashed cut-line is shown in the bottom panel. The profile is shifted by a constant value. \textbf{e)} Similar to  \textbf{(d)} but under a positive $V_{SD}$. Gradient vectors (white arrows) of the potential indicating the direction of electron flow show how the electron drifts towards the island. Modulation of the CB energy profile along the cut-line with the positive $V_{SD}$ is shown in the bottom panel. Reset due to the drain field is shown as the black arrow.}
\end{figure}

Figure \ref{fig2}a illustrates the steps involved in the detection cycle, namely, 1) photo absorption by the BP; 2) ultra-fast electron transfer into MoS\2 using the built-in field, followed by directing the photo-electron to an electron trapping island in the MoS\2 layer; 3) sensing the single electron through a resistance fluctuation in the FET, and finally, 4) detrapping of the electron and resetting the detector. We discuss each step in detail below, with the paragraphs following same numbering as the above points.

\paragraph{1. Absorption enhancement:}
Efficient absorption of the incoming photon is critical in achieving good external quantum efficiency (EQE). Linear dichroism for low energy transition\cite{Deng18} arising from the optical selection rules results in the absorption of $1550~nm$ photons polarized along zig-zag direction to be at least two orders of magnitude less than that along the armchair direction\citep{Yuan15,Schuster15}. Although this gives the SPD polarization sensitivity, absorption of unploarized light caps at $50\%$. We use unpolarized light to characterize the device for measurements detailed later in the text. \txb{Other absorption pathways such as inter-layer absorption is negligible and the light absorption in the device is through absorption in the direct bandgap of BP.}

The thickness of the absorbing region in this work is $\unsim 40$ times lower than the typical values for SPADs\cite{Vines19}. With the help of the interference effects\cite{Wang17,Huang19} produced by the back metal reflector (gate contact) and by optimizing the top and bottom hBN thicknesses, we enhance the BP absorption at $1550~nm$ and compensate for the ultra-thin nature of the BP layer. The simulation results from a  transfer matrix method (TMM)\cite{Ansari18} of the propagation of the waves using complex refractive indices from literature\cite{Lee19, Schuster15, Ermolaev20, Munkhbat22, Olmon12, Mash73, Gao12, Shkondin17} are shown in Figure \ref{fig2}b, and verified using further experiments (Supplementary Note S1). We fabricate devices considering these, and the estimated absorption at $1550~nm$ by the devices for unpolarized light (light polarized along armchair) is $25.9\%~(51.8\%)$ for device A and $26.85\%~(53.7\%)$ for device B. Thus, without having to resort to external resonators or similar structures\cite{Li16}, we can achieve good absorption.

\paragraph{2. Photocarrier separation and capture:}
To achieve a good external quantum efficiency (EQE), we need to separate the photo electrons before they recombine with holes in p-type BP. The ultra-thin nature of the BP layer and the type II band alignment (Figures \ref{fig2}a and \ref{fig2}c) help to separate the e-h pair through vertical transport in ultra-fast time scales\cite{Hong14,Murali19}. Once the electron reaches the depleted MoS\2 region, it is confined vertically by the conduction band (CB) offsets and can be transported efficiently in the X-Y plane.

To understand the potential landscape of the device and how the various screening factors affect it, we simulate a 3D model of our device. The conduction band energy of the MoS\2 layer along the X-Y plane for zero source to drain voltage ($V_{SD}=0~V$), and $V_G<0~V$ is given in Figure \ref{fig2}d (top panel). The regions of MoS\2 that overlap with the other layers in the device are annotated. The simulated profile indicates a potential energy dip (dark blue regions) at the region where the Gr overlaps with the WSe\2/MoS\2/BP region and is also shown in the energy profile along the orange dashed cut-line shown in the bottom panel. This is a result of screening of the gate field by the WSe\2 and Gr layers and due to the presence of the p-type BP region. The biasing condition ($V_{SD}>0~V$), generates a lateral electric field directed from the source end to the drain end, assisting the photo-electrons in drifting toward the low energy region near the source end. The gradient vectors in the potential shown as white arrows (Figure \ref{fig2}e) indicate the trajectory of the electrons. Once the electron reaches this region, they experience confinement in the lateral directions due to the surrounding high-energy regions. Plus, the already existing vertical confinement results in the formation of an island for trapping electrons locally inside MoS\2. It should be noted that the regions where BP, MoS2 and drain graphene overlap needs to be minimized to loss of photoelectrons before they are directed to the island.

The immediate vertical transfer of electrons from BP to MoS\2 and the highly efficient transport of electrons in MoS\2 to the island owing to the careful design, enables us to realize devices with a large active area. The fabricated devices have active areas of $\unsim 135~\mu m^2$ for device A and $\unsim 245~\mu m^2$ for device B. Such large active areas are critical for efficient light coupling.

\paragraph{3. Sensing single electron capture:}
To generate a detectable signal from the capture of a single electron, we need to couple it with a high-gain mechanism. In the proposed scheme, this gain is a result of two effects. First, we enhance the carrier density fluctuation created by the single photo-generated electron. The small volume of the island helps in this. For example, for the thickness of MoS\2 used in this work ($\unsim5~nm$), having an island of lateral dimensions of $1~\mu m \times 1~\mu m$ will result in a change in the local carrier density in MoS\2 by \txb{$\frac{1~electron}{1~\mu m \times 1~\mu m \times 5~nm}=2\times 10^{14}~cm^{-3} $} on the addition of a single electron. This fluctuation is large as the MoS\2 is operated in a depleted condition (negative $V_G$) and is the primary factor in achieving a high gain. The band alignment of the stack before trapping an electron is shown as dashed lines in Figure \ref{fig2}c. When an electron occupies the island, the potential in its vicinity changes. WSe\2 and MoS\2 having low carrier concentrations, experience a strong modulation as shown by solid lines in Figure \ref{fig2}c. This modulation is enhanced by the close proximity of the local island in the MoS\2, with the WSe\2 region. \txb{Reducing the thickness of WSe\2 could enhance this modulation. On the other hand, a very thin WSe\2 could result in loss of the trapped electron by tunneling into source-Gr. Hence an optimum thickness of WSe\2 is required.}

From the careful design, the island and the source region of the FET are self-aligned. The screening of the gate electric field by the Gr\cite{Silkin21} at the sourcing end and the lack of such screening at the channel results in the drain current ($I_D$) to be limited by the source injection efficiency, the effects of which can be seen as a saturating trend in $I_D$ towards more negative $V_G$ (Supplementary Figure S2b). The FET resistance is now tightly correlated with the charging state of the island through coulomb interactions. When a photoelectron occupies the island, the resultant photogating lowers the barrier seen by the holes in source-Gr to transfer into the WSe\2 channel or to tunnel into the BP channel. The large number of holes experiencing this modulation provide further gain and result in a reduction in FET resistance. The holes injected into WSe\2 or tunneled into BP can then laterally drift across the channel. \txb{From our measurements on a device with no overlap between BP and drain-Gr (Supplementary Figure S8), we believe that the tunneling into the BP is the dominant path for holes. This tunneling can be enhanced by choosing a thinner MoS\2.} Although the source-Gr can have a larger overlap with the WSe\2 flake, the majority of the hole injection occurs towards the edge of the graphene flake due to current crowding effects\cite{English16} and is thus the region with the most sensitivity.

\paragraph{4. Resetting the device:}
Once the signal is read out, the electron in the island needs to be taken out to avoid charge build-up and maintain high sensitivity. We realize this with the help of the source to drain field. The same field that was used to direct the electron to the island in MoS\2 also helps to pull the electron out of the island and be collected at the source-Gr as shown by a black arrow in Figure \ref{fig2}a and in the bottom panel of Figure \ref{fig2}e. This results in the hole injection efficiency restoring back to its previous lower value and thus results in an increment in the overall device resistance, marking the auto reset of the detector.

\subsection{Measurement and SPD performance evaluation}
\begin{figure}[H]
  \centering
  \includegraphics[width=16.5cm]{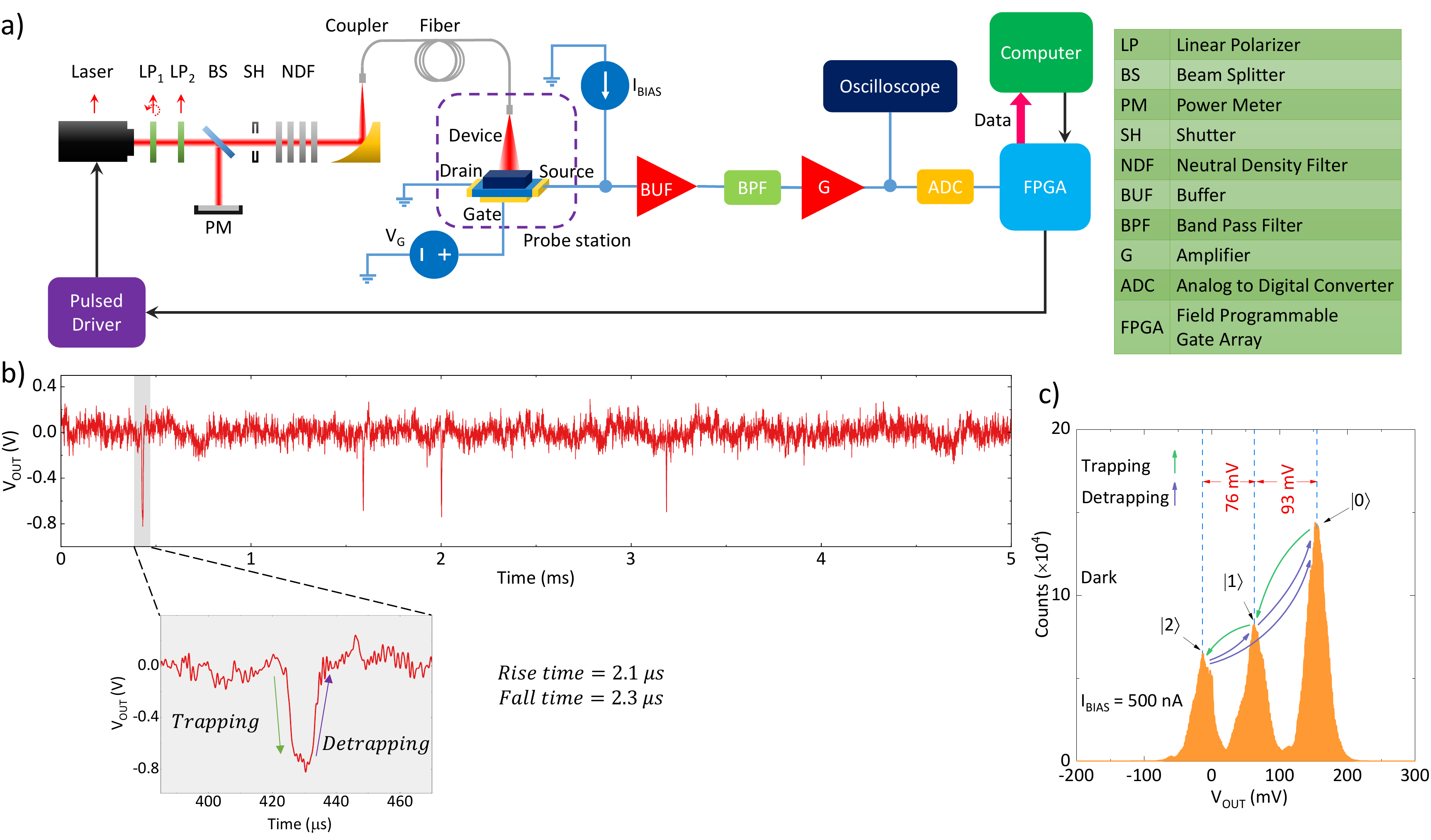}	
  \caption{\label{fig3}\textbf{Experimental setup and device response.} \textbf{a)} Measurement setup used to generate the low occupation number photon states with Poisson distribution needed for evaluating the SPD and the associated electronics for data processing and control. The red arrows indicate the polarization of the beam at that element. Laser and LP$_2$ have matched linear polarization, whereas LP$_1$ is rotated. Black arrows indicate control signals. \textbf{b)} A representative time trace of $V_{OUT}$ collected at $300~K$ and under dark conditions. The DC offset is removed by the BPF and the signal is amplified. The downward spikes indicate the electron detection events. A zoomed in view of a typical spike is shown at the bottom. The electron trapping results in a sudden drop in $V_{OUT}$ as seen in the figure. Due to the auto reset, the electron quickly detraps resulting in the recovery of $V_{OUT}$ to the previous state. The device exhibits a $2.3~\mu s$ fall time and a $2.1~\mu s$ rise time. \textbf{c)} Measured histogram of $V_{OUT}$ collected at $300~K$ and a $V_G=-4~V$, under dark conditions showing distinct peaks corresponding to different electron occupation states in the island. Green (violet) arrows indicate the trapping (de-trapping) events. }
\end{figure}

To make statistically reliable predictions, we use the measurement setup shown in Figure \ref{fig3}a and acquire enough data ($\unsim 3.5$ million detection events). \txb{Images of the experiment setup used are shown in Supplementary Figure S9.} It is critical to accurately calibrate the photon flux to make reliable predictions. A heavily attenuated laser is commonly recommended for photon source preparation\citep{Coldenstrodt09,garrison08,migdall13,Lundeen09}. This rules out many of the uncertainties that can arise if we use other sources such as a light emitting diode. We follow this procedure and verify the measured standard deviations in all the stages of photon source preparation stays well below the uncertainty of the power meter ($\pm5\%$).
More details of the laser positioning, detailed calibration steps and other stages in the setup can be found in \textbf{Methods}. To monitor transient FET resistance fluctuations, we pass a fixed current, $I_{BIAS}$, through the source-Gr into the channel, bias the gate at $V_G=-4V$, and measure the resultant voltage generated at the source terminal ($V_{OUT}$). Capture (release) of an electron in the island will thus result in a reduction (increase) of the $V_{OUT}$. A representative time trace of the $V_{OUT}$ collected at room temperature under dark conditions using a high bandwidth setup is shown in Figure \ref{fig3}b, exhibiting multiple downward spikes in $V_{OUT}$. A zoomed-in view of a typical spike is shown at the bottom, showing the electron capture (falling edge) and electron release (rising edge). We achieve a fall time of $2.3~\mu s$ and a rise time of $2.1~\mu s$. Since the processes involved are fast in nature, we expect the intrinsic speed of the detector to be even better with increasing bandwidth of the measurement setup or resorting to on-chip processesing circuitry as used in Si or InGaAs SPADs.

The clear distinct peaks in the histogram of $V_{OUT}$ under dark conditions plotted in Figure \ref{fig3}c correspond to the different occupation numbers in the island. We label the rightmost peak as $\ket{0}$ corresponding to zero electrons occupying the island. When an electron is trapped, the occupation number increases by one, and $V_{OUT}$ drops to a lower value. The peaks are also separated by comparable changes in $V_{OUT}$, hinting at the ability of the device to detect new electrons to some extent even before the previously trapped electron is pumped out of the island. Thus, unlike SPADs, the detector is not completely dead, following a detection event.

The high sensitivity of SPDs makes them prone to generate false-positive signals. The most common of these are dark counts, which is the main reason InGaAs SPADs require cooling to temperatures $\unsim230~K$ to operate\cite{migdall13}. In our design, the origin of dark counts at room temperature can be attributed to the injection of stray electrons, the thermal generation of electrons in any of the layers in the system, or from absorption of black body radiated photons. We do observe some device-to-device variation in the dark counts owing to the variabilities introduced by the manual fabrication process. The minimum observed dark count is from device A ($\unsim 720~Hz$), which compares favorably against commercial InGaAs SPADs\cite{com1,com2,com3} (few kHz in free running mode at 230 K). If required, the dark counts can be lowered further as shown through their rapid decrease with reducing temperature (Supplementary Figure S10).

Any variation in the dark count rate is also a strong indicator of degradation in the heterostructure quality. For the duration of the study, and also with thermal cycling from 7K to 300K, we didn't observe any increase in the dark count of the device (Supplementary Figure S11) as a result of the excellent protection from the hBN encapsulation (followed by vacuum annealing) and the robustness of the heterostructure.

For characterization, the dark condition is realized by blocking the pulsed laser output with a shutter (Figure \ref{fig3}a). The count rates didn't change between the measurement conditions (shutter closed, laser ON), and (shutter open, laser OFF) ruling out the presence of any stray photon incidence paths. To extract the efficiency, we then keep the laser ON, and measure the difference in counts with the shutter open, and closed for varying the average number of photons per pulse ($\bar{n}$). A sample time trace collected at (shutter open, laser ON) with many peaks exhibiting high correlation to laser firing instants is shown in Supplementary Figure S12. The left axis of Figure \ref{fig4}a shows the rise in photon detection events with an increasing number of photons per laser pulse for device A (operated under an $I_{BIAS}$ of $2~\mu A$) and is comparable to the count rates of commercial free running InGaAs SPADs configured for reasonable dark counts. The photon flux is obtained by multiplying $\bar{n}$ with the pulse repetition frequency, $f$. The extracted EQE (fraction of detection events from all the photons impinging on the device) saturates at $19.6\%$. The measured EQE is maximum for low values of $\bar{n}$. The photon counting setup has a lower bandwidth (as we have separately verified) and is expected to be the reason for the saturation of the count rates and the drop in efficiency at photon flux above $\unsim20~kHz$. To demonstrate repeatability of the observations, we measure another sample, device B (more details in Supplementary Note S2). Here the analysis was performed for varying $\bar{n}$ as well as varying $f$. For each pulse repetition frequency, $\bar{n}$ was maintained between $0.0053$ and $0.2652$ to ensure that the probability of more than one photon falling on the device is negligible. The results for different pulse repetition frequencies are shown in Figure \ref{fig4}b. At low photon flux, device B exhibited a similar EQE of $21.4\%$. The high EQE also indicates a high internal quantum efficiency (fraction of detection events from the absorbed photons) of $75\%$ for device A and $79\%$ for device B. We estimate the EQE at a fixed $\bar{n}$, as a function of $f$ (red stars in Figure \ref{fig4}c). The linear fit (blue trace) matches the experimental data well. The slope of the line gives the product of EQE and $\bar{n}$, when $\bar{n}$ is sufficiently low. We extract an EQE of $20.2\%$ from this analysis, closely matching the previously estimated EQE. We also measure similar EQE ($\unsim20\%$) from another fabricated device. Aligining the polarization of all the impinging photons with the armchair direction of BP should double this EQE. Even without correcting for the polarization anisotropy, the measured efficiencies of these devices are comparable to that of commercial InGaAs/InP APDs ($10-30\%$)\cite{com1,com2,com3} operating at temperatures of $\unsim 230~K$.

\section{Conclusions}
\begin{table}
\renewcommand{\arraystretch}{1.5} 
\begin{center}
\begin{tabular}{ |m{2.2cm}| m{3.3cm} | R{0.9cm}| R{1cm} | R{1.0cm} | R{1.8cm} | m{3.0cm} | } 
\hline
\multicolumn{1}{|>{\centering}m{2.2cm}|}{\textbf{System}} &
\multicolumn{1}{>{\centering}m{3.3cm}|}{\textbf{Principle}} &
\multicolumn{1}{>{\centering}m{0.9cm}|}{\textbf{$\lambda$ (nm)}} &
\multicolumn{1}{>{\centering}m{1.0cm}|}{\textbf{$\eta$ (\%)}} &
\multicolumn{1}{>{\centering}m{1.0cm}|}{\textbf{T (K)}} &
\multicolumn{1}{>{\centering}m{1.8cm}|}{\textbf{DCR}} &
\multicolumn{1}{>{\centering}m{3.0cm}|}{\textbf{Speed}}\\
\hline
\txr{\textbf{This Work}}  &	\txr{\textbf{Photo Gating}}    &	\txr{\textbf{1550}}   &	\txr{\textbf{21}}    &	\txr{\textbf{300}}   &	\txr{\textbf{720 Hz}} &	\txr{\textbf{\mbox{rise $< 2.1~\micro s$,} \mbox{fall $< 2.3~\micro s$}}}\\\hline
Gr JJ\cite{Evan21} & Super Conducting & 1550& 13 & 0.027 &  & \\\hline
InAs QD\cite{Li07} & Photo Gating & 1310 & 0.35 & 4.5 & 1.58 kHz& \\\hline
InGaAs GaAs NW\cite{Farrell19} & Avalanche & 1064 & Low & 77 & 10 Hz & 	\\\hline
InGaAs QD + GaAs\cite{Gansen07} & Photo Gating & 805 &  & 4  &  & \\\hline
InAs QD + AlGaAs\cite{Kardynal07} & Photo Gating & 684 & 1.3 & 4.2 & & 200ns rise\\\hline
InAs QD\cite{Shields00} & Photo Gating & Red & 0.48 & 4	 &  & 	\\\hline
MoS\2/BLG\cite{Roy18}    &	Photo Gating    &	609    &	4    &	80    &	0.07 Hz&\\\hline	
InAs QD + GaAs\cite{Blakesley05}    &   Photo Gating  &   550 &   12.5    &   5   &   2 mHz    &	$\sim$100 s (repopulation of QD)\\\hline
CdS NW\cite{Luo18}  &	Photo Gating    &	457   &	23    &	300   &	1.87 mHz &	rise 4ms, recovery 100s\\\hline
\end{tabular}
\end{center}
    \caption{\textbf{SPDs based on low dimensional materials}: Benchmarking with existing literature on SPDs based on low dimensional materials. $\lambda-$ Operating wavelength, $\eta-$ efficiency, $T-$ operating temperature, $DCR-$ dark count rate.}
    \label{tab:T1}
\end{table}

Even though photodetectors based on nanomaterials have been extensively studied in the past, their use in single-photon detectors is very limited. To emphasize the technological advancements put forward by the proposed scheme, we benchmark our detector with reports from existing literature\cite{Evan21,Li07,Farrell19,Gansen07,Kardynal07,Shields00,Roy18,Blakesley05,Luo18} on single-photon detection based on low dimensional systems. The results tabulated in Table \ref{tab:T1} show the pressing need for near infrared SPDs capable of room temperature operation. To the best of our knowledge, this is the first report of SPD built using nano-materials capable of detecting single photons at $1550~nm$ while maintaining room temperature operation.

To conclude, we designed a system that can detect single-electron fluctuation and use it to realize a single-photon detector operating at room temperature. Our design's flexible nature also provides the opportunity to quickly adapt the proposed idea for the detection of single photons at higher wavelengths beyond the capabilities of InGaAs based detectors. This prototype demonstration of an alternate approach towards single photon detection could be augmented with many of the advancements made over the last few decades in the fields of Si and InGaAs SPDs such as active gating, high bandwidth readout circuitry, etc. Additional resources unique to this design such as the ability of the island to hold multiple electrons and the anisotropic absorption in BP could also be leveraged to attain photon number resolution and polarization sensitivity. Future research into how we can utilize these existing and new resources is needed to fully utilize the potential of the proposed single photon detection scheme. Nevertheless, the good efficiency at the telecom wavelength of $1550~nm$ and the large active area demonstrated by the fabricated devices already make the proposed design a strong candidate for future single-photon detection applications.

\section{Methods}
\subsection{Fabrication}
We start with a Si wafer with $285~nm$ oxide grown by dry chlorinated thermal oxidation and forming gas annealing. Pre-patterned contacts for the source, drain, and gate regions are defined by optical lithography using a $360~nm$ UV source and a positive photoresist (AZ5214E). A $30~nm$ thick Ti layer followed by a $40~nm$ thick Au layer was deposited by DC magnetron sputtering and was lifted off using an acetone/isopropyl alcohol rinse to form the pre-patterned contacts. Flakes are exfoliated from a bulk crystal and are transferred to a poly dimethyl-siloxane (PDMS) stamp using Nitto tape. An hBN flake of the required thickness is identified by optical contrast. It is then dry transferred under an optical microscope (with precise position and rotation control) to the gate contact to form the bottom gate dielectric. This is followed by the transfer of two few-layer graphene flakes contacting the source and drain contacts following the same method. A $\unsim 5~nm$ thick WSe\2 flake is then transferred on top, contacting the two graphene regions. MoS\2 and BP flakes are then transferred on top to form the quantum well. It is ensured that the BP flake remains isolated from the Gr and WSe\2 regions by the MoS\2 flake to avoid stray current paths. It is also critical to ensure that the area of the vertical source-Gr/WSe\2/MoS\2/BP junction is optimized. The stack is then capped with a top hBN layer of suitable thickness to protect the BP from the environment. The device is then annealed at $200~^\circ C$ for 3 hours under pressure lower than $10^{-6}$ Torr.

\subsection{DC characterization and data acquisition}
The device is loaded into a DC probe station (Lakeshore CRX-6.5K) with sealed viewports and is maintained at a vacuum of $<10^{-4}$ Torr. DC characterization of the device is done using a parameter analyzer (Keithley 4200A SCS). For SPD measurements, with the generated weak coherent states coupled, the device is biased with the necessary $I_{BIAS}$ and $V_G$, and the generated $V_{OUT}$ is observed. A unity gain buffer (Texas Instruments LMP7721MAEVALMF/NOPB) and a fixed gain amplifier (Texas Instruments THS3091EVM) are added to the signal path for impedance matching and amplification, respectively. The signal is DC offset as required to maintain the input in the operating range of the buffer. The amplified signal is then fed to a digital storage oscilloscope (Tektronix MDO3000). The signal is also fed to a bandpass filter to filter out external electrical noise. The filtered signal is then fed to a high-speed 14-bit ADC (Digilent Zmod Scope 1410) interfaced with a field programmable gate array (FPGA) combined with a processing system (Digilent Eclypse Z7). The FPGA does the required thresholding of the signal and the fast counting, and the results are communicated to a computer through the onboard processing system. The measurement setup is illustrated in Figure 3a.

\subsection{Generation of weak low occupation number photon states}
The generation of pure single-photon states is typically done by processes such as single-photon parametric down-conversion. Such processes offer very low flexibility, and hence, weak laser pulses are typically employed in characterizing single-photon detectors$^{39}$. When operated well above the threshold, states emitted by a laser can be approximated by coherent states of light that exhibit a Poisson probability distribution for the different constituent Fock states$^{40}$. On attenuation, coherent states retain this nature and only experience a reduction in the mean photon number ($\bar{n}=|\alpha|^2$)$^{39}$. To generate these coherent states, we start with a laser operated well above the threshold and lasing at $1550~nm$. The laser is driven by a pulsed driver, which is controlled by the FPGA. Software executed in a computer communicates with the FPGA and sets the properties of the laser pulses. The generated laser pulses pass through two linear polarizers (LP$_1$ and LP$_2$) [see Figure 3a]. The polarization axis of LP$_2$ is precisely aligned with the laser polarization. LP$_1$ is rotated by different angles to control the intensity, while $LP_2$ is kept at a fixed angle parallel to the laser polarization, making sure that the output polarization is the same for the varying intensity. This is the tuning knob that is used to control $\bar{n}$ for the final states. A portion of the beam is sampled using a beam splitter (BS). A power meter (PM) is attached to one arm of the BS to monitor the power. The exact measurement of power is critical in estimating efficiency. Hence, we use a PM with NIST traceable calibration (Thorlabs S132C). The beam exiting the other arm of BS passes through a mechanical shutter (SH), followed by multiple absorbing anti-reflection coated attenuators (NDF). This heavily attenuated laser pulse is then coupled to the device through a multi-mode optical fiber. Detailed procedure$^{41-42}$ of calibrating the measurement setup can be found in \textbf{Supplementay Note S3}. As the fiber is not polarization-maintaining and is multi-mode, the high mode mixing results in the light exiting the fiber being unpolarized. This is verified by analyzing the polarization of the light output by the fiber (Supplementary Figure S14). The standard deviation in the measured power is well below the measurement uncertainty of the power meter ($\pm5\%$) across the range of powers used. The dark counts remained the same as when the shutter was open, and the laser was off. Hence, for further measurements, the laser was kept firing photons, and only the state of the shutter was toggled between light and dark measurements. The probe states of different $\bar{n}$ are then used to characterize the device, and the device response is accumulated for a $\unsim 3.5$ million events in total to estimate the maximum efficiency.


\subsection{Coherent states}
A coherent state $\ket{\alpha}$ can be represented in the Fock state basis as
\begin{equation}
\ket{\alpha}=e^{-\frac{|\alpha|^2}{2}}\sum_{n=0}^{\infty} \frac{\alpha^n}{\sqrt{n!}}\ket{n}
\end{equation}
where $\ket{n}$ are the photon basis states with $\ket{0}$ representing the vacuum or the zero photon state. Thus, the probability of the coherent state $\ket{\alpha}$ containing $n$ photons is given by
\begin{equation}
\begin{split}
P(n)&=|\braket{n}{\alpha}|^2\\
&=e^{-|\alpha|^2}\frac{|\alpha|^{2n}}{n!}\\
&=e^{-\bar{n}}\frac{{\bar{n}}^n}{n!}
\end{split}
\end{equation}
with the mean photon number $\bar{n}=|\alpha|^2$ that can be related to the measured average optical power $\overline{P}$ as
\begin{equation}
\overline{P}=\bar{n} h \nu f
\end{equation}
$h$ the Planck's constant, $\nu$ frequency of incoming photon and $f$ the repetition rate of the laser.

\section{Data availability}
All data needed to evaluate the conclusions in the paper are present in the paper and/or in the Supplementary Materials. Additional data related to this paper may be requested from the corresponding author.

\section{Acknowledgements}
This work was supported in part by a Core Research Grant from the Science and Engineering Research Board (SERB) under Department of Science and Technology (DST), a grant from Indian Space Research Organization (ISRO), a grant from MHRD under STARS, and a grant from MHRD, MeitY and DST Nano Mission through NNetRA. K.W. and T.T. acknowledge support from the JSPS KAKENHI (Grant Numbers 21H05233 and 23H02052) and World Premier International Research Center Initiative (WPI), MEXT, Japan.

\section{Conflict of interest}
The authors declare no financial or non-financial conflict of interest.

\section{Contributions}
N.A. and K.M. conceived the idea, analyzed the data and co-wrote the manuscript. N.A. performed the measurements. K.W. and T.T. supplied materials for the work.


\bibliography{ref.bib}
%
\newpage
\begin{figure}[H]
  \centering
  \includegraphics[width=15.9cm]{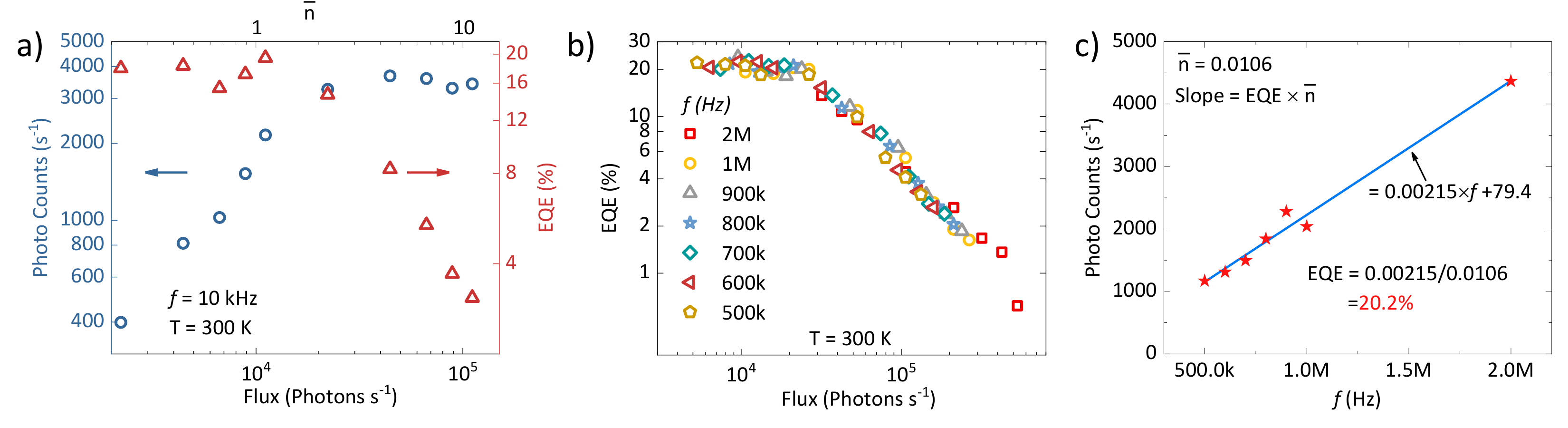}	
  \caption{\label{fig4}\textbf{SPD efficiency estimation.} \textbf{a)} Measured counts (dark count subtracted) under an $I_{BIAS} = 2~\mu A$ as a function of increasing photon flux on the left axis and the estimated EQE for unpolarized light detection on the right axis for device A. \textbf{b)} Repeatability of the observations by measuring the single-photon detection efficiency in device B operated under $I_{BIAS}=2~\mu A$ at different trigger frequencies. A doubling of EQE will be observed for polarized photon detection, and an IQE of $79\%$ is estimated. c) Variation of the photo-counts as a function of laser repetition frequency (red stars) and a linear fit (blue trace). The EQE estimated from the slope closely matches the efficiency of $21.4$ measured using other methods. The low offset $c=79.4$ also points to the low error in the measurement.}
\end{figure}
\newpage
\AtEndDocument{

\section*{Supplementary material}

Table S1: **Existing SPDs for 1550 nm operation**: A representative summary of the existing SPD technologies for operation at 1550 nm. $\lambda-$ Operating wavelength, $\eta-$ efficiency, $T-$ operating temperature, $DCR-$ dark count rate.

| System | Principle | $\lambda$ (nm) | $\eta$ (%) | T (K) | DCR | Speed |
|---|---|---|---|---|---|---|
| TES[9] | Super Conducting | 1556 | 95 | <0.1 | | 800 ns recovery |
| SNSPD[6] | Super Conducting | 1550 | 95 | 2.1 | 100 Hz | 42 ns recovery |
| InGaAs/InP[19] | Avalanche | 1550 | 40 | 213 | 2.3 kHz | 10 $\mu s$ hold off |

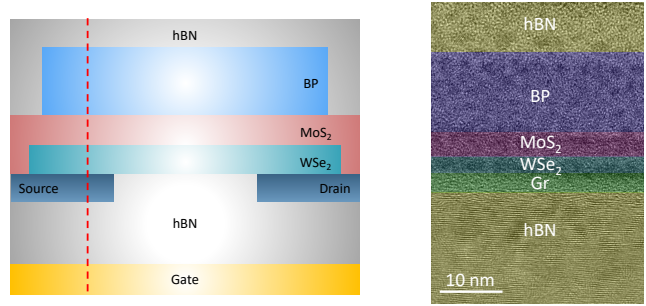

Figure S1: **Device cross-section.** False-coloured HR-TEM of the device (right panel) along the red dashed line in the left panel. The acceleration voltage was set to 300 keV

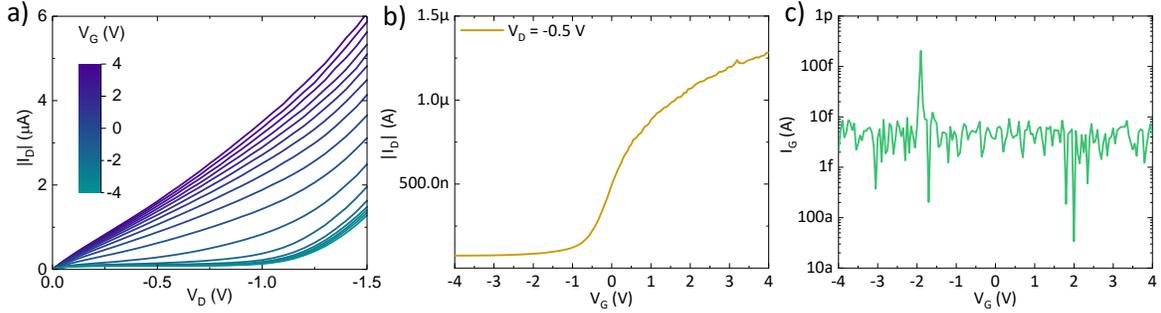

Figure S2: **DC characteristics.** **a)** Modulation of $I_D - V_D$ characteristics of the device as a function of $V_G$ collected at 300 $K$. For negative $V_G$, the plot shows the behavior when holes are injected from the source contact. **b)** Modulation of $I_D$ as a function of $V_G$ for a fixed bias of $V_D = -0.5$ $V$. At a positive $V_G$, the WSe$_2$, MoS$_2$ and BP regions are n-type doped, and the transport is dominated by electrons. On applying a negative $V_G$, the MoS$_2$ channel turns OFF, and the current is dominated by the hole transport by tunneling into the p-type BP and possibly through lateral conduction across p-type WSe$_2$. The gate field screening by the hole injecting graphene results in saturation of $I_D$ for negative $V_G$. In the device, the parallel MoS$_2$ and the WSe$_2$ channels and the hole path through tunneling into BP reduces the overall ON/OFF ratio of the FET, as shown in (a). **c)** Gate leakage current $I_G$ measured from the device shows an ultra-low leakage limited by the measuring setup verifying the absence of any dielectric degradation.

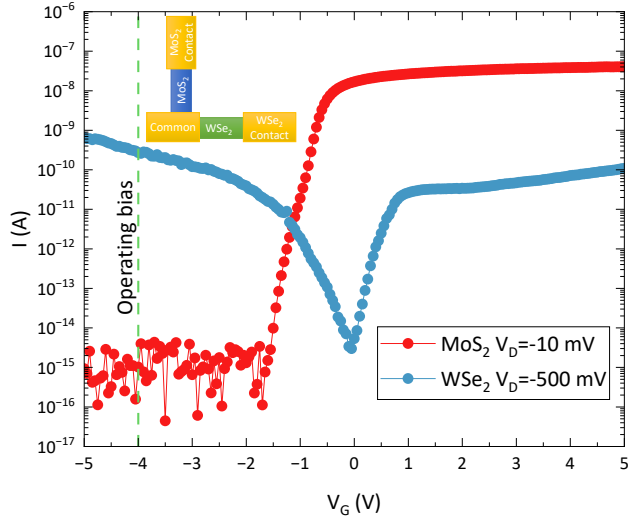

Figure S3: **FET characteristics.** Modulation of current through $MoS_2$ layer and $WSe_2$ layer as a function of $V_G$ from a separate experiment. The flakes were of similar thickness as in the SPD device. Note that the $MoS_2$ channel is completely depleted of electrons for $V_G < -1.7V$. Biasing the SPD at $V_G = -4$ ensures that the FET current is entirely due to hole transport by tunneling into BP and possibly laterally across the $WSe_2$ layer.

## S1. Absorption estimation

The absorption estimation in the main text is obtained from TMM simulation. In order to validate these observations, we compare the results predicted from a TMM simulation with that from an experiment. As the fabricated SPDs have contacts made of few-layer graphene very close to the active area, we fabricate a separate stack without graphene contacts for validating our TMM simulation. This will avoid errors that can arise from the absorption of photons by graphene while determining the absorption in the BP layer. The stack contains (top to bottom) hBN (18 nm), BP (13 nm), $MoS_2$ (6 nm), hBN (54.2 nm), Au (42 nm), Ti (31 nm), $SiO_2$ (285 nm) and Si. The thickness of the flakes were determined using atomic force microscopy (AFM) and that of metals using cross-sectional TEM. Because of the presence of metal layers, the transmittance is negligible, and we use a reflection-based method to estimate absorption. The schematic of the setup is shown in Supplementary Figure S4. The reflected power is recorded for various polarization angles ($\theta$) by rotating

the half-wave plate by $\theta/2$. The ratio of the reflected power on the full stack versus that on top-hBN/bottom-hBN/Au/Ti/SiO$_2$/Si is presented in Supplementary Figure S5 and shows reasonable agreement with the simulations from the TMM. This validates the predictions from our TMM simulation. The complex refractive indices used here are obtained from the literature [22,27-33]. The values at $\lambda = 1550\ nm$ are given in Supplementary Table S2.

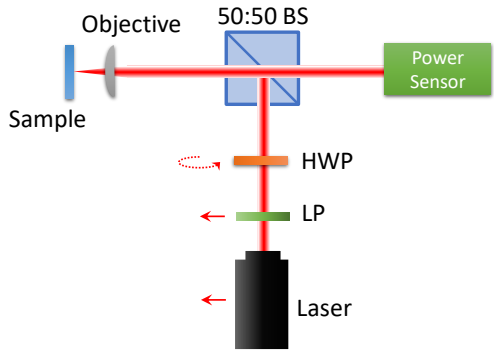

Figure S4: **Absorption measurement setup.** Setup used to estimate absorption from the reflected power. $LP-$ linear polarizer, $HWP-$ half-wave plate, $BS-$ beamsplitter.

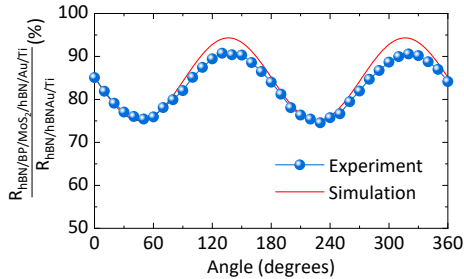

Figure S5: **Relection measurement.** Angle dependent ratio of the reflected power from the top-hBN/BP/MoS$_2$/bottom-hBN/Au/Ti/SiO$_2$/Si stack and top-hBN/bottom-hBN/Au/Ti/SiO$_2$/Si (blue circles) showing reasonable agreement with the predictions from TMM simulation (red trace). Ratio peaks when the incident polarization is aligned with the zig-zag direction of BP.

To extract the absolute value of absorption, we need to account for all the losses in the measured power. In this discussion, *'metal film'* refers to the Au/Ti/SiO$_2$/Si structure, and *'stack'* refers to the full hBN/BP/MoS$_2$/hBN/Au/Ti/SiO$_2$/Si structure. Few assumptions have to be made in order to estimate the absorption. First, we assume 100% reflection from *metal film*. This is reasonable owing to the high reflectivity of Au and is further supported

Table S2: **Complex refractive index of various materials at 1550 nm**: Values of n and k at 1550 nm for the different materials extracted from literature and used in TMM.

| Ref | Material | n | k |
|---|---|---|---|
|  | Air | 1 | 0 |
| 29 | hBN | 2.0737 | 0 |
| 24 | BP | 3.8056 | 0.3013 |
| 30 | MoS$_2$ | 4.1645 | 0 |
| 31 | WSe$_2$ | 3.9949 | 0 |
| 32 | Au | 0.3226 | 10.62 |
| 33 | Ti | 3.5983 | 4.0483 |
| 34 | SiO$_2$ | 1.4658 | 0 |
| 35 | Si | 3.487 | 0 |

by a reflection of 97.6% predicted from TMM for this structure. For a power, $P^0$ incident on the *metal film*, $r_M P^0$ is reflected back. $r_M$ is the fraction of light reflected back from *metal film*. Due to the losses in the system, the power measured on the power sensor $P_M$ is given by $P_M = r_M P^0 \times L$, where $L$ captures the loss in the setup. From the assumption of $r_M \approx 1$, and measuring $P^0$ and $P_M$, we can estimate $L$. We assume $L$ remains constant throughout the measurement. For an arbitrary direction $\phi$, we can write the equation,

$$\begin{aligned} P^0 &= r_{Stack}^\phi P^0 + a_{Stack}^\phi P^0 \\ &= \frac{P_{Stack}^\phi}{L} + a_{Stack}^\phi P^0 \end{aligned} \quad (1)$$

where $r_{Stack}^\phi$ is the fraction of light reflected by the *stack*, and $a_{Stack}^\phi$ is the fraction of light absorbed by the *stack* for the light polarization defined by $\phi$, and $P_{Stack}^\phi$ is the measured reflected power.

For the special case of zig-zag (ZZ) direction, we can write,

$$P^0 = \frac{P^{ZZ}_{Stack}}{L} + a^{ZZ}_{Stack}P^0 \qquad (2)$$

where $P^{ZZ}_{Stack}$ is the reflected power measured, and $a^{ZZ}_{Stack}$ is the fraction of light absorbed by the *stack* when light is polarized along ZZ. From equations 1 and 2, we get,

$$a^{\phi}_{Stack} - a^{ZZ}_{Stack} = \frac{P^{ZZ}_{Stack} - P^{\phi}_{Stack}}{P^0 \times L} \qquad (3)$$

On the *stack*, the hBN, BP, and MoS$_2$ layers create interference effects which is the prime reason for observing absorption enhancement in BP. This effect also increases the absorption in Au and Ti and needs to be considered in absorption calculation. For the low thickness of BP, the absorption by the Au and Ti layers while in the *stack* shows negligible modulation with light polarization, which gets significant as the BP thickness increase. For the thickness of BP used in this *stack*, TMM results predict a minimum absorption of 8.09% for light polarized along armchair (AC) and a maximum of 8.64% for ZZ by the Au and Ti films, which are reasonably close. Hence, equation 3 is a good approximation for the absorption by the BP layer while in the stack. Results from the experiment and from the simulation are given in Supplementary Figure S6. The simulated value of absorption is obtained by integrating the position-dependent absorption (shown in the middle panel of Supplementary Figure S7) over the BP region.

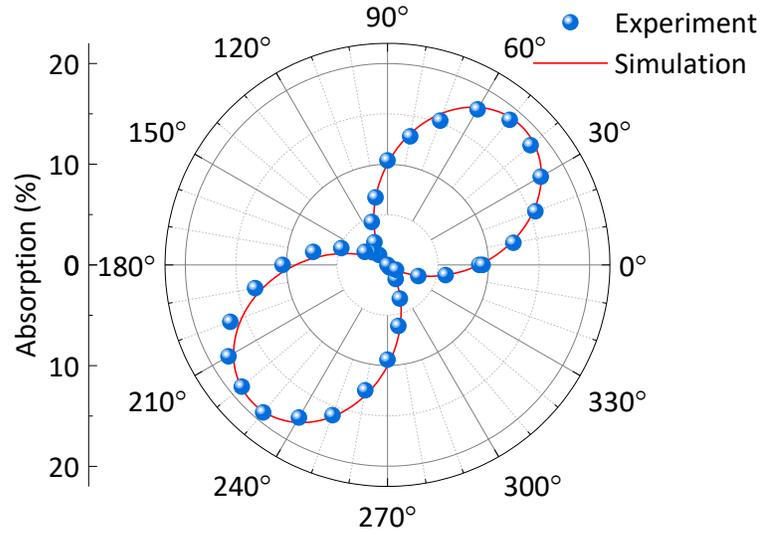

Figure S6: **Estimated absorption.** Absorption estimated from experimental data using equation 3 (blue circles) and the absorption predicted from TMM in polar plot. Absorption shows two clear lobes as expected, and the angle at which the absorption peaks corresponds to the incident light polarization matching with the armchair direction in BP.

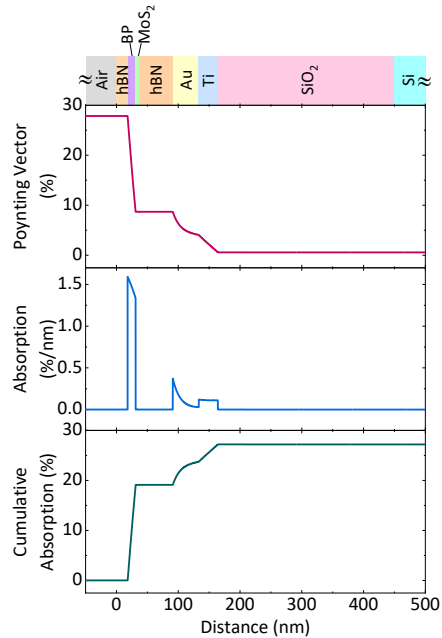

Figure S7: **Transfer Matrix Method simulation.** Results from TMM simulation showing the position-dependent Poynting vector as a fraction of the incident (top panel), absorption/nm (middle panel) showing a large absorption where the BP layer is present, and the integrated absorption (bottom panel).

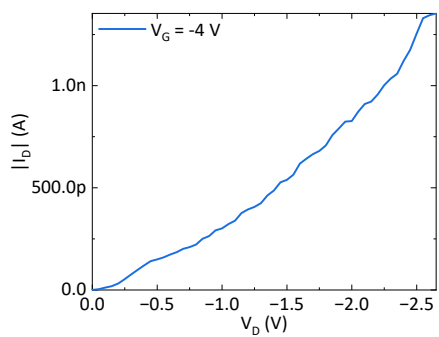

Figure S8: **Hole current path.** Measured $I_D$ vs $V_D$ at $V_G = -4$ V for a device with BP having no overlap with drain-Gr. The low current compared to Figure S2a suggest the transport through BP is the dominant hole transport pathway.

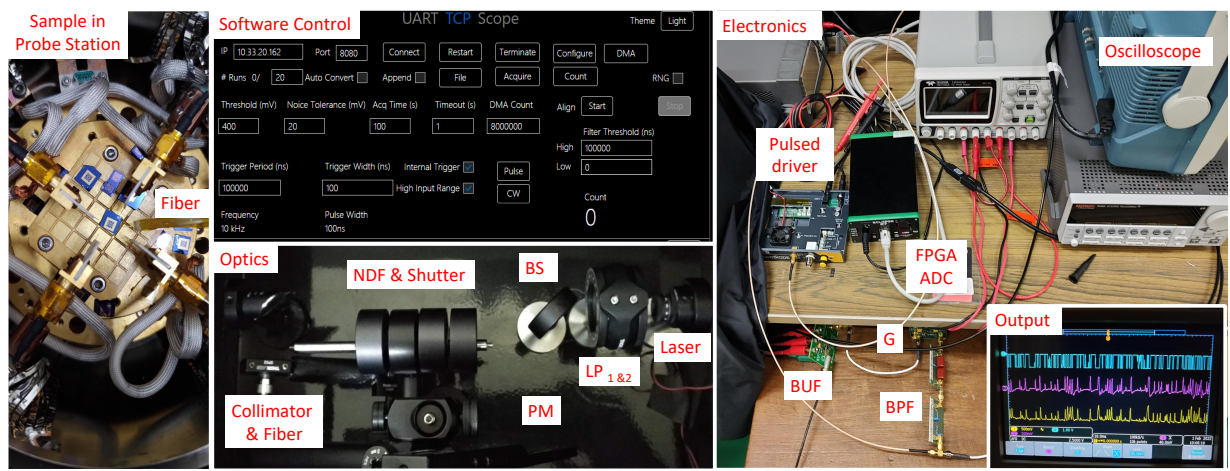

Figure S9: **Experiment setup.** Left: Sample loaded inside the probestation where single photon measurements were carried out. Middle top: Developed software interface for communicating with FPGA. Middle bottom: Optics for photon flux tuning and fiber launching. Right: Circuits for signal amplification, filtering, FPGA, oscilloscope and pulsed driver for laser. Right bottom: Various output signals from FPGA for sanity checks.

9

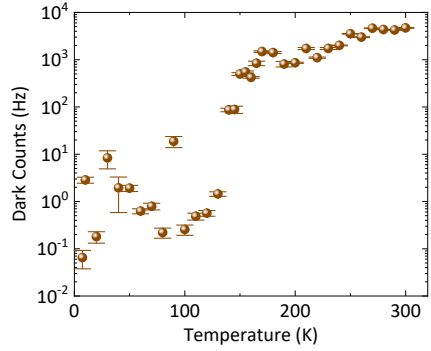

Figure S10: **Temperature dependence of dark counts.** Measured dark counts as a function of temperature. Error bars indicate the standard error. The very low error bar verifies that the statistical fluctuations are small in our measurements.

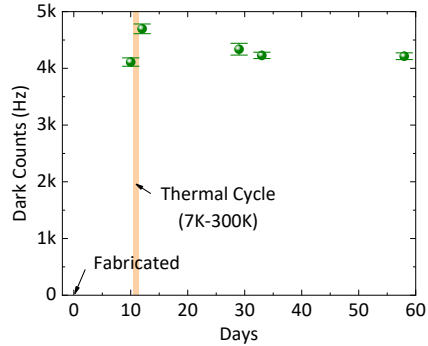

Figure S11: **Device stability.** Measured dark counts at 300K showing no increase with days, verifying the stability of the device and the good protection offered by the hBN encapsulation. The device didn't exhibit any degradation even with thermal cycling from 300K to 7K and back to 300K.

## S2. Repeatability of the detector

Another device, B was fabricated and measured at 300K to verify the repeatability of the observations. The thickness of all the layers were maintained similar to device A to reduce uncertainties. Device B exhibited a resistance ∼1.5 time lower than device A, owing to variability in the fabrication. The active area was measured to be ∼245 $\mu m^2$, compared to ∼135 $\mu m^2$ for device A. The efficiency at a fixed laser repetition frequency was then measured as a function of photon flux. Device B exhibited a maximum EQE of 21.4%. This corresponds to an IQE of 79%, comparable to the IQE of 75% for device A. We did not

10

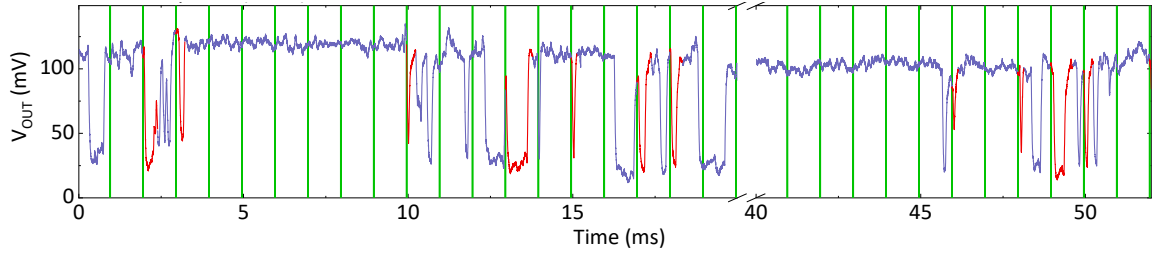

Figure S12: **Time trace with the shutter open**. Green vertical lines indicate the time instants of laser firing. Under the experimental conditions, each firing event has a ∼69.7% probability of containing photons with correct polarization that can be absorbed by BP. The detection events occurring immediately after laser firing (green vertical lines) are highlighted in red, heralding the detection of a photon. The detection events shown in blue, which have no correlation to laser firing is the dark counts of the device.

observe significant difference in IQE among these devices of different active areas, indicating that the processes, including the collection of carriers are very efficient.

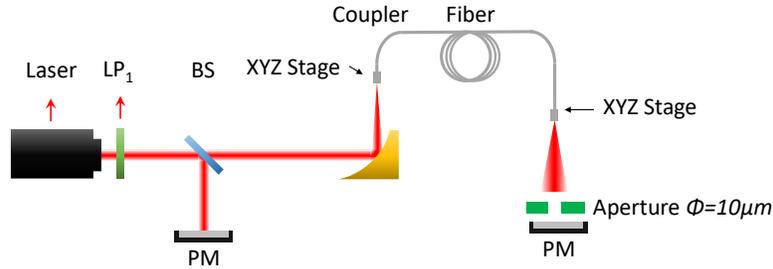

Figure S13: **Source calibration.** The ratio of power after the aperture with that from the beamsplitter's second output port and the calibrated NDF attenuation is used to estimate the photon flux.

## S3. Photon source calibration

Accurate calibration of the photon source is critical to extract the performance of the detector-under-test-reliably. We follow the experiment setup discussed in [41,42] to create photon states with variable average photon numbers. The polarizer $LP_1$ is aligned with the laser polarization. The beam is split using a non-polarizing beamsplitter. The beam from the first output port of the beamsplitter is launched into a multimode fiber using an off-axis parabolic mirror to focus the beam. The fiber tip is mounted on a micro-positioning

11

multi-axis stage to ensure maximum coupling as shown in Supplementary Figure S13. The ratio between the power exiting the fiber and the power from the second output port of the beamsplitter is obtained using a commercial germanium power sensor (Thorlabs S132C). The neutral density filters (NDFs) are placed on flip mounts to easily switch into and out of the beam path before the parabolic mirror. The NDFs are kept stable using alignment rods to ensure negligible beam deflection. High-quality antireflection coating on both sides of the NDF helps in minimizing errors. The attenuation of each of the NDFs is verified individually by measuring the total power exiting with and without the NDF in the beam path. The measured attenuation of the NDF falls within the manufacturer's specifications. To estimate the power falling on the device, an aperture (10 $\mu m$ diameter) of size smaller than the active area of our devices was mounted on the power meter. The other end of the fiber (the end from which the beam exits) is positioned perpendicular to the aperture plane. It is moved using a micro-positioning multi-axis stage till maximum power is recorded on the power meter (corresponding to the aperture being aligned with the beam center). The ratio between the power exiting the aperture and the power measured at the second output port is calculated. This ratio is then multiplied by the ratio between device active area and aperture area. The aperture being smaller than the device area guarantees that we are not underestimating the power falling on the device. By knowing this ratio and the attenuation of each NDF, we can reliably estimate the upper limit of the power falling on the device. During measurements, the fiber tip is again positioned with the device using a micro-positioning multi-axis stage to give the maximum response. With the upper limit of power falling on the device known, we can reliably estimate the mean photon number in each pulse, and then the lower limit of the device efficiency.

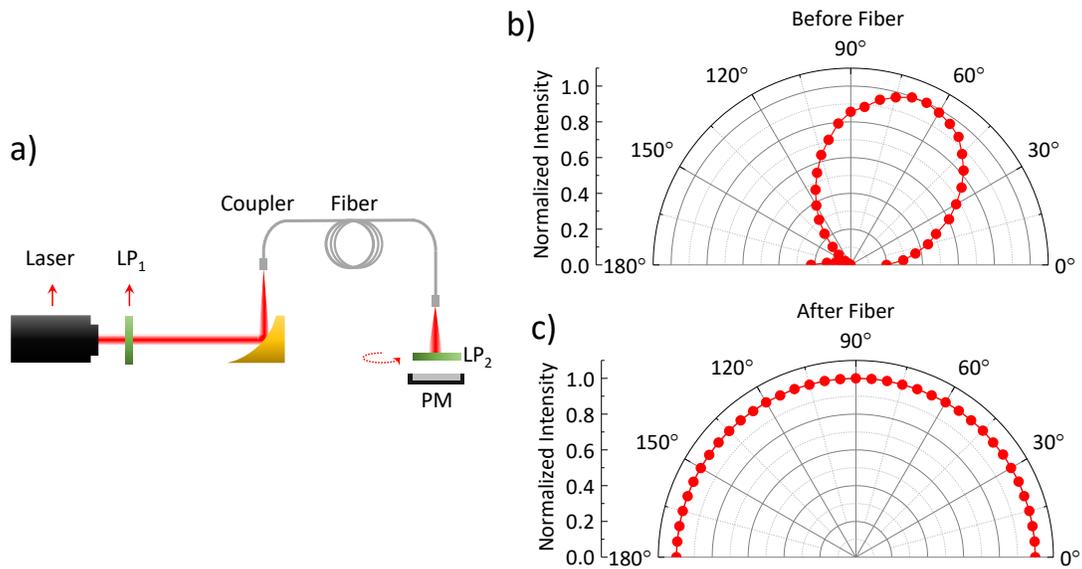

Figure S14: **Polarization property of the beam.** a) Setup used to measure the polarization of the beam. Linear polarizer, $LP_1$ is aligned with the laser polarization. b) Results from the power measurements obtained as a function of the $LP_2$ rotation angle when $LP_2$ and power meter (PM) was kept immediately after $LP_1$. c) Similar measurements as in (b), but with $LP_2$ and PM placed after the fiber. The strong mode mixing in the multi-mode fiber results in a complete loss of polarization. During the measurement, care was taken not to cause any disturbance leading to a stress-induced change in the mode mixing in the fiber. We also didn't observe any significant fluctuation in power measured at each angle with time.

}
\end{document}